\documentstyle[prl,aps,twocolumn,psfig]{revtex}

\begin{document}
\preprint{LBNL-39742}
\twocolumn[\hsize\textwidth\columnwidth\hsize\csname @twocolumnfalse\endcsname
\title{Study Medium-induced Parton Energy Loss in $\gamma$+jet Events
  of High-Energy Heavy-Ion Collisions}
\author{Xin-Nian Wang}
\address{Nuclear Science Division, MS 70A-3307, \\
\baselineskip=12pt Lawrence Berkeley National Laboratory, Berkeley, CA 94720}
\author{Zheng Huang}
\address{Department of Physics, University of Arizona, Tucson, AZ 85721}

\date{January 6, 1997}
\maketitle

\begin{abstract}

The effect of medium-induced parton energy loss on 
jet fragmentation is studied in high-energy heavy-ion collisions.
It is shown that an effective jet fragmentation function can be 
extracted from the inclusive $p_T$ spectrum of charged particles
in the opposite direction of a tagged direct photon with
a fixed transverse energy. We study the modification of the
effective jet fragmentation function due to parton energy
loss in $AA$ as compared to $pp$ collisions, including $E_T$
smearing from initial state radiations for the photon-tagged jets.
The effective fragmentation function at $z=p_T/E_T^\gamma\sim 1$
in $pA$ collisions is shown to be sensitive to the additional
$E_T$ smearing due to initial multiple parton scatterings
whose effect must be subtracted out in $AA$ collisions in order
to extract the effective parton energy loss. Jet quenching
in deeply inelastic lepton-nucleus scatterings as a measure
of the parton energy loss in cold nuclear matter is also
discussed. We also comment on the experimental feasibilities
of the proposed study at the RHIC and LHC energies and some
alternative measurements such as using $Z^0$ as a tag at the
LHC energy.
\end{abstract}
]


\section{Introduction}
Hard processes are considered good tools to study ultrarelativistic
heavy-ion collisions because they happen early in the reaction
processes and thus can probe the early stage of the evolution of
a dense system, during which a quark-gluon plasma (QGP) could 
exist for a short period of time. Among the proposed hard
probes, large transverse momentum jets or partons are especially
useful because they interact strongly with the medium. For example,
an enhanced acoplanarity and energy imbalance of two back-to-back
jets \cite{aco} due to multiple scatterings, jet quenching due to
the medium-induced radiative energy loss of a high-energy parton
propagating through a dense medium \cite{qn1} can provide important 
information on the properties of the medium and interaction
processes that may lead to partial thermalization of the produced
parton system. The medium-induced radiative energy loss of
a fast parton traversing a dense QCD medium is also interesting 
by itself because it illustrates the importance of quantum
interference effects in QCD. As recent studies have
demonstrated \cite{lpm1,lpm2}, it is very important to
take into account the destructive interference among
many different radiation amplitudes induced by multiple scatterings
in the calculation of the final radiation spectrum.
The so-called Landau-Pomeranchuk-Midgal effect \cite{lpm0}
can lead to very interesting, and sometimes nonintuitive
results for the radiative energy loss of a fast parton inside
a QCD medium. Recently, Baier, Dokshitzer, Mueller, Peign\'e 
and Schiff (BDMPS)  showed \cite{lpm2} that the energy loss per
unit distance, $dE/dx$, grows linearly with the total length
of the medium, $L$,  which in turn can be related to the total
transverse momentum broadening squared, $\Delta k_T^2$, of
the parton from multiple scatterings. It turns out that
both quantities, $dE/dx$ and $\Delta k_T^2$, are related
to the parton density of the medium that the parton is
traveling through. One can therefore determine the parton
density of the produced dense matter by measuring the energy
loss of a fast parton in high-energy heavy-ion collisions.

Unlike in the QED case, where one can measure directly the radiative
photon spectrum and thus the energy loss of a fast electron, one
cannot measure directly the energy loss of a fast parton in QCD.
Since a parton is experimentally associated with a jet, a cluster
of hadrons in a finite region of the phase space, 
an identified jet can contain
particles both from the fragmentation of the leading parton
and from the radiated partons. If we neglect the $k_T$ broadening
effect, the total hadronic energy contained in a jet should not
change even if the leading parton suffers radiative energy loss.
However, significant changes could happen to particle distributions
inside the jet, or the fragmentation function and jet profile,
due to the induced radiation of the leading parton. Therefore,
one can measure the radiative energy loss indirectly via the
modification of the jet fragmentation function and jet profile.

A jet fragmentation function is defined as the particle distribution
in the fractional energy. In order to measure the fragmentation function
one has to first determine the initial energy of the fragmenting parton
either through other measured kinematic variables as in $e^+e^-$ and
$e^-p$ or calorimetric measurements as in $pp$ and $p\bar{p}$ collisions.
However, because of the large value of $dE_T/dyd\phi$ and its fluctuation
in high-energy heavy-ion collisions \cite{wg90}, the conventional
calorimetric study of a jet cannot determine the jet energy to
such an accuracy as required to determine the parton energy loss. In 
search for an alternative measurement, Wang and Gyulassy \cite{qn2}
proposed that single-particle $p_T$ spectrum can be used to study the
effect of parton energy loss, since the suppression of large $E_T$
partons naturally leads to the suppression of large $p_T$ particles.
Because the particle spectrum at large $p_T$ is the
convolution of jet cross sections
and fragmentation functions, particles with a fixed $p_T$
can come from the fragmentation of partons of different initial
energies with some average value $\langle E_T^{\rm jet}\rangle$. 
In this case, the suppression of the $p_T$ spectrum due to parton 
energy loss is then related to the modification of jet fragmentation 
function at an averaged $\langle z\rangle=p_T/\langle E_T^{\rm jet}\rangle$. 
Since $\langle E_T^{\rm jet}\rangle$ is approximately proportional 
to $p_T$, by varying $p_T$ one can then study the energy dependence of 
the modification of jet fragmentation functions at a fixed
$\langle z\rangle$ \cite{wang2}.

In order to study the modification of the {\it whole} jet fragmentation
function due to parton energy loss in the full $z$ range, we and 
Sarcevic \cite{whs} proposed to measure the particle $p_T$ distribution 
in the opposite transverse direction of a tagged direct photon.
Since a direct photon in the central rapidity region ($y=0$) is
always accompanied by a jet in the opposite transverse direction
with roughly equal transverse energy, the $p_T$ distribution
of particles in that direction is directly related to the
jet fragmentation function with known initial energy, 
$E_T^{\rm jet}\approx E_T^{\gamma}$. In such $\gamma + {\rm jet}$
events, the background due to particle production from the
rest of the system was estimated to be well below the
$p_T$ spectrum from jet fragmentation at moderate large $p_T$. 
Therefore, one can easily extract the fragmentation function
from the experimental data without much statistical errors
introduced by the subtraction of the background. By comparing
the extracted jet fragmentation function in $AA$ to that in $pp$
collisions, one can then measure the modification of
the fragmentation function and determine the parton energy loss.

Because of the complexity of the problem, it will be helpful for
us to first discuss all possible relevant processes which might
have some effects on the study of parton energy loss in $\gamma+{\rm jet}$
events in high-energy $AA$ collisions. One can order the processes
in a chronological order with respect to the hard process of
direct photon production.

(1) As the two nuclei approach and pass
through each other, the two participating beam partons which 
later produce the direct photon will suffer initial state interactions 
with other oncoming nucleons. The participating beam partons will
then suffer radiative energy loss and acquire transverse momentum kicks
because of these soft interactions. The initial state interactions
will also cause the shadowing of the parton distributions inside
the nuclei. These initial state effects will certainly affect the 
production rate of direct photons at a given transverse energy $E_T^\gamma$.
However, they will not influence the propagation and fragmentation
of the accompanying produced jet parton in events triggered with
a direct photon with a fixed $E_T^\gamma$.

(2) After the hard process
in which a direct photon and a jet parton are produced, the jet
parton can also scatter from the beam nucleons within a tube of
a transverse size at most 1 fm in the rest part of the colliding nuclei
which has not passed through. Since the colliding nuclei pass 
through each within a very short period of time, $t\sim 1$ fm/$c$
(this is the spatial size of wee-partons, while the valence
partons have a Lorentz contracted size of $R_A m_N^2/2s$), the produced 
jet parton in central rapidity region will not have time to 
interact with other beam nucleons outside the tube. 
Since we are only interested in jet partons in
the central rapidity region, these scatterings will not cause
the jet partons to lose transverse energy. Rather, together with
the initial state interactions, they will change the final
transverse momentum of the jet parton, resulting in an $E_T$ broadening
in addition to the $E_T$-smearing caused by initial state radiations
associated with the hard process.
The effects of this $E_T$ broadening should also exist and can be 
studied independently in $pA$ collisions. 

(3) In the triggered events,
there are many other processes which can also produce hard or
semihard partons \cite{report} and thus form a dense medium
in the central rapidity region. The photon-tagged jet parton will 
then interact with these partons during its propagation through 
the dense medium. We call these interactions as final state interactions.
The induced radiative energy loss and transverse momentum broadening,
referred to as $k_T$-broadening in this paper, are the focus of our study.

In this paper, we will study in detail the effect of parton energy loss
on the jet fragmentation function as extracted from the
$p_T$ spectrum in the opposite direction of a triggered
direct photon. In particular, we will take into account
the $E_T$ smearing of the jet due to initial state radiations
associated with the $\gamma + {\rm jet}$ processes. 
We will show that the particle spectrum from the jet fragmentation 
at $p_T\sim E_T^\gamma$ is very sensitive to the $E_T$ broadening
from initial and final state scatterings with beam partons. 
One can then use our proposed
measurement to determine the $E_T$ broadening in $pA$
collisions. This small but finite effect must then be subtracted out
when one determines the medium-induced parton energy loss in $AA$ collisions.
We will also investigate the sensitivity of the modification of the
fragmentation function to the energy and $A$ dependence of the parton 
energy loss. The change of the profile function in the azimuthal
angle due to the $k_T$-broadening of the parton from multiple
scatterings inside the medium will also be discussed. Here
$k_T$ is the parton transverse momentum with respect to the
original jet direction, which can be related to the parton energy
loss according to BDMPS study \cite{lpm2}. 
Finally, we will discuss the experimental feasibility of the 
proposed study and alternative measurements using $Z^0$ particles as a tag. 
We will also discuss how similar measurements can be made in 
deeply inelastic lepton-nucleus scatterings, from 
which one can determine the energy loss of a fast parton 
passing through a cold nuclear matter.

\section{Modified Fragmentation Functions}

The fragmentation functions of partons hadronizing in the vacuum 
have been studied extensively in $e^+e^-$, $ep$ and $p\bar{p}$ 
collisions \cite{mattig}. These functions describe particle 
distributions in the fractional energy, $z=E_h/E_{jet}$, 
in the direction of a jet. Similar to parton distributions inside
hadrons, the fragmentation functions are also nonperturbative
in nature. However, parton cascades during the early stage of
the fragmentation can be described by perturbative QCD.
The measured dependence of the fragmentation functions 
on the momentum scale is shown to satisfy the QCD evolution
equations very well. We will use the parametrizations of the most
recent analysis \cite{bkk} in both $z$ and $Q^2$ dependence for 
jet fragmentation functions $D^0_{h/a}(z,Q^2)$ to describe the 
fragmentation of a parton ($a$) into hadrons ($h$) in the  vacuum.

The fragmentation of a parton inside a medium is different from
that in the vacuum, because of its final state interactions with
the medium and the associated radiations. Such interactions
and medium-induced radiations will cause the deflection 
and energy loss of the propagating parton which in effect
will modify the fragmentation functions from their corresponding forms
in the vacuum. In principle, one could study the modification 
of jet fragmentation functions in perturbative QCD
in which induced radiation of a propagating parton in a medium and 
Landau-Pomeranchuk-Migdal interference effect can be dynamically
taken into account. However, for the purpose of our current
study, we can use a phenomenological model to describe the
modification of the jet fragmentation function due to an
effective energy loss $dE/dx$ of the parton. Such an approach
is useful and possibly necessary for experimental studies
of the parton energy loss and multiple final state scatterings.

In this phenomenological model we assume that the size and
life time of the system is small compared to the hadronization
time of a fast parton. In the case of a QGP, a parton cannot
hadronize inside the deconfined phase. A fast parton will
hadronize outside the system and the fragmentation can be 
described as in $e^+e^-$ collisions, however, with reduced parton energy. 
The interaction of a parton $a$ with the medium can be characterized
by the mean-free-path $\lambda_a$ of parton scatterings, the radiative
energy loss per scattering $\epsilon_a$ and the transverse momentum
broadening squared $\Delta k_T^2$. The energy loss per unit distance
is thus $dE_a/dx=\epsilon_a/\lambda_a$ which in principle depends on
$\Delta k_T^2$ and $\lambda_a$. We assume that the probability 
for a parton to scatter $n$ times within distance $L$ is given 
by a Poisson distribution,
\begin{equation}
  P_a(n,L)=\frac{(L/\lambda_a)^n}{n!} 
  e ^{-L/\lambda_a} \; .
\end{equation}
We also assume that the mean-free-path of a gluon is half
that of a quark, and the energy loss $dE/dx$ is twice that
of a quark. The emitted gluons, each carring energy $\epsilon_a$ 
on the average, are assumed to hadronize also according to the fragmentation 
function. For simplification, we will neglect the energy fluctuation
given by the radiation spectrum for the emitted gluons. We assume 
the momentum scale in the fragmentation function for the emitted 
gluons to be set by the minimum scale $Q_0^2= 2.0 $ GeV$^2$.
Since the emitted gluons only produce hadrons with very small 
fractional energy, the final modified fragmentation functions 
in the moderately large $z$ region are not very sensitive to 
the actual radiation spectrum and the momentum scale dependence 
of the fragmentation functions for the emitted gluons.
In this paper, we will also neglect possible final state 
interactions between hadrons from parton fragmentation and the 
hadronic environment at the late stage of the evolution of the
whole system. However, it is important for future investigations
to estimate the influence of pure hadron scatterings on the final
observed jet fragmentation functions.

We will consider parton fragmentation in the central rapidity 
region of high-energy heavy-ion collisions. In this case, we only
need to study partons with initial transverse energy $E_T$ and 
traveling in the transverse direction in a cylindrical system. 
With the above assumptions, the modified fragmentation functions 
for a parton traveling a total distance $L$ can be 
approximated as \cite{whs},
\begin{eqnarray}
  D_{h/a}(z,L,Q^2)& =&
  \frac{1}{C^a_N}\sum_{n=0}^NP_a(n,L)\frac{z^a_n}{z}D^0_{h/a}(z^a_n,Q^2)
  \nonumber \\
  &+&\langle n_a\rangle\frac{z'_a}{z}D^0_{h/g}(z'_a,Q_0^2), 
  \label{eq:frg1}
\end{eqnarray}
where $z^a_n=z/(1-n\epsilon_a/E_T)$, $z'_a=zE_T/\epsilon_a$ and
$C^a_N=\sum_{n=0}^N P_a(n)$. $D^0_{h/a}(z,Q^2)$ are the jet 
fragmentation functions in the vacuum which we take the 
parametrized form in Ref.~\cite{bkk}. We limit the number of 
inelastic scatterings to $N=E_T/\epsilon_a$ to conserve momentum.
One can check that the above modified fragmentation functions
satisfy the momentum sum rule by construction,
$\sum_h \int  z D_{h/a}(z,L,Q^2)dz
=\sum_h \int z D^0_{h/a}(z,Q^2) dz=1$.  
{}For large values of $N$, the average number of scatterings
within a distance $L$ is approximately 
$\langle n_a\rangle \approx L/\lambda_a$. The first term 
in the above equation corresponds to the fragmentation of 
the leading partons with reduced energy $E_T-n\epsilon_a$
after $n$ inelastic scatterings. This term normally dominates
for leading particles in the moderate and large $z$ region. 
Since the fragmentation functions $D^0_{h/a}(z,Q^2)$ generally 
decrease with $z$, especially quite rapidly at moderate and 
large $z$ region, the reduction in energy will lead to the
suppression of leading particles or the decrease of the
fragmentation functions in this region as compared to
the case in vacuum. The second term in the above equation
comes from the emitted gluons each having energy $\epsilon_a$ on 
the average. This term is generally significant only in the
small $z$ region and it increases the effective fragmentation
functions in the small $z$ region, or enhances soft particle
production. We should note that our assumptions on the hadronization
of the emitted gluons are too schematic to give a quantitative
description of the physics involved in that small $z$ region.

For a given parton energy $E_T$ and the total distance $L$, 
the above effective fragmentation functions depend on only two
parameters, the mean-free-path $\lambda_a$ and energy loss
per scattering $\epsilon_a$. As demonstrated in Ref.~\cite{whs},
contributions from the leading partons who have suffered at least
one inelastic scattering is completely suppressed for $z$ values 
close to 1. The remaining contribution comes from those partons
which escape the system without a single inelastic scattering,
with a probability $\exp(-L/\lambda_a)$, which depends
on $\lambda_a$ but is independent of the parton energy $E_T$ and 
the parton energy loss $dE_a/dx$. On the other hand, in the
intermediate $z$ region, particles from the fragmentation of
the leading partons with reduced energy dominates. The suppression
of the fragmentation functions is controlled by the total
energy loss, $\langle\Delta E_T\rangle=\langle n_a\rangle\epsilon_a
=L dE_a/dx$, which depends only on $dE_a/dx$.
One, therefore, could determine in principle these two parameters,
$\lambda_a$ and $dE_a/dx$, simultaneously from the measured 
suppression of the effective fragmentation functions, for 
fixed $E_T$ and $L$. However, as we will see in the next
section, the complication of not knowing the jet energy precisely
will render such arguments unrealistic. In certain cases, one has 
to resort to a model-dependent global fitting of the modification 
of the fragmentation functions in order to determine the mean-free-path
and parton energy loss.

\section{The Inclusive Fragmentation Function of Photon-tagged Jets}

As we have emphasized in the Introduction, the most important point
in the study of parton energy loss through the measurement
of the modification of the parton fragmentation functions
is the determination of the initial parton or jet energy. However,
the direct measurement of a jet energy to the accuracy as required
to determine an energy loss of a few GeV is unfeasible due to the
large background and its fluctuation in high-energy heavy-ion
collisions. To overcome this difficulty, it was proposed \cite{whs}
that direct photons in heavy-ion collisions can be used to tag
the energy of jets which always accompany the direct photons.
Because of initial state radiations associated with the production
of a direct photon, the accompanying jet is not always exactly
in the opposite direction of the photon and its transverse
energy also differs from collision to collision, though the
averaged jet energy is well approximated by the energy of the
triggered photon. In Ref.~\cite{whs}, the variation of jet energy
was not considered in the study of the effective inclusive jet fragmentation
function in $\gamma+{\rm jet}$ events and the modification due
to parton energy loss. In this paper, however, we would like to explore
the effect of $E_T$ smearing due to initial state radiations.
We will see that such a smearing complicates the simple
procedure to determine the parton energy loss and the mean-free-path
of final state scatterings as outlined in the previous study \cite{whs}.

Let us consider events which have a direct photon with fixed transverse 
energy $E^{\gamma}_T$ in the central rapidity region, 
$|y_\gamma|\leq \Delta y/2$, $\Delta y=1$.
Given the jet fragmentation functions $D^0_{h/a}(z)$, with $z$ 
the fractions of momenta of the jet carried by hadrons, one can 
calculate the differential $p_T$ distribution of hadrons,
averaged over the kinematical region $(\Delta y,\Delta \phi)$,
from the fragmentation of a photon-tagged jet in $pp$ collisions,
\begin{eqnarray}
  \frac{dN_{pp}^{\gamma-h^\pm}}{dyd^2p_T}&\equiv&
  \frac{1}{d\sigma^{\gamma}_{pp}/dy_\gamma dE_T^\gamma}
  \frac{d\sigma^{\gamma-h^\pm}_{pp}}{dy_\gamma dE_T^\gamma dyd^2p_T}
  \nonumber \\
  \frac{d\sigma^{\gamma-h^\pm}_{pp}}{dy_\gamma dE_T^\gamma dyd^2p_T}&=&
  \sum_{a,h}\int dE_T^a dy_a d\phi_a
  \frac{d\sigma^{\gamma-a}_{pp}}{dE_T^\gamma
   dy_\gamma dE_T^a dy_ad\phi_a}   \nonumber \\
  & &\hspace{-1.0in}\times \frac{D_{h/a}^0(p_T/E_T^a)}{p_TE_T^a}
  \int_{(\Delta y,\Delta\phi)}\frac{dy}{\Delta y}\frac{d\phi}{\Delta\phi}
  f_0(y_a-y,\phi_a-\phi) \label{eq:crs-g}
\end{eqnarray}
where the summation is over jet ($a$) and hadron ($h$)
species and $f_0(y,\phi)$, 
assumed to be the same for all hadron species, is the normalized 
hadron intrinsic profile around the parton axis.
If the azimuthal angle of the photon is $\phi_{\gamma}$ and 
$\bar{\phi}_{\gamma}=\phi_{\gamma}+\pi$, the restricted kinematical
region for the selected hadrons is defined as
$(\Delta y,\Delta \phi)=(|y|\leq \Delta y/2, |\phi-\bar{\phi}_{\gamma}|
\leq \Delta\phi/2)$. One could also use more complicated
geometry, such as a circle with a given radius, for the phase
space restriction to define a jet. The inclusive differential 
cross section for direct photon production is
\begin{equation}
  \frac{d\sigma^{\gamma}_{pp}}{dy_\gamma dE_T^\gamma}
  =\sum_a \int dE_T^a dy_a d\phi_a 
  \frac{d\sigma^{\gamma-a}_{pp}}{dE_T^\gamma dy_\gamma dE_T^a dy_ad\phi_a}
  \;\; .
\end{equation}

We now define the $E_T$-smearing function, $g_{pp}(E_T^a.E_T^\gamma)$,
and parton correlation function, $f_{\rm jet}(y_a,\phi_a)$, as
\begin{equation}
  g_{pp}(E_T^a,E_T^\gamma)f_{\rm jet}(y_a,\phi_a)
  \equiv \frac{1}{\frac{d\sigma^{\gamma}_{pp}}{dy_\gamma dE_T^\gamma}}
  \frac{d\sigma^{\gamma-a}_{pp}}{dE_T^\gamma
   dy_\gamma dE_T^a dy_ad\phi_a}. \label{eq:smear0}
\end{equation}

In a perturbative calculation to the lowest order in $\alpha_s$,
which was used in our earlier study \cite{whs}, the $E_T$-smearing
function and parton correlation function are simply two $\delta$-functions, 
$g_{pp}(E_T^a,E_T^\gamma)=\delta(E_T^a-E_T^\gamma)$,
$f_{\rm jet}(y_a,\phi_a)\propto\delta(\phi_a-\bar{\phi}_{\gamma})$, due to
momentum conservation. The intrinsic hadron profile function $f_0(y,\phi)$
in this leading order calculation should be the measured jet profile.
In calculations beyond the leading order, the photon and jet parton have 
a finite imbalance in transverse momentum due to the initial state 
radiations. The final state radiations also contribute to the measured 
jet profile $f(y,\phi)$ which should be the convolution of the parton
correlation function $f_{\rm jet}(y_a,\phi_a)$ from perturbative 
calculations to a given order and the intrinsic hadron 
profile $f_0(y,\phi)$,
\begin{equation}
  f(y,\phi)=\int dy_a d\phi_a f_{\rm jet}(y_a,\phi_a)
  f_0(y_a-y,\phi_a-\phi)\;\; .
\end{equation}

Note that the differential cross section  $d\sigma^{\gamma-a}$
and the fragmentation function $D^0_{h/a}(z,Q^2)$ in Eq.~(\ref{eq:crs-g})
both depend on the factorization scheme and the associated scale.
So are the $E_T$-smearing function and parton correlation function in
Eq.~(\ref{eq:smear0}). Unlike the collinear divergences in
the next-to-leading order jet cross sections, which are cancelled via the
definition of a jet with a finite size, 
the collinear divergences in Eq.~(\ref{eq:crs-g})
are subtracted out via the definitions of parton distributions and 
fragmentation functions. Therefore, the differential cross 
section $d\sigma^{\gamma-a}$ beyond the 
leading order in Eq.~(\ref{eq:crs-g}) is not the same as the
$\gamma-{\rm jet}$ cross section in which a jet is defined via
the transverse energy within a finite region in phase space \cite{soper}.
The two cross sections can be related via some divergence-free physical 
observables, {e.g.}, total hadronic energy within the 
acceptance $(\Delta y,\Delta \phi)$, which can be computed from 
Eq.~(\ref{eq:crs-g}).

With the above definitions of the $E_T$-smearing function
and jet profile function, we can rewrite Eq.~(\ref{eq:crs-g}) as
\begin{eqnarray}
  \frac{dN_{pp}^{\gamma-h^\pm}}{dyd^2p_T}&=&\sum_{a,h}r_a(E_T^{\gamma})
  \int dE_T g_{pp}(E_T,E_T^{\gamma}) \nonumber \\
  &\times & \frac{D_{h/a}^0(p_T/E_T)}{p_T E_T} 
  \frac{C(\Delta y,\Delta\phi)}{\Delta y\Delta\phi}, \label{eq:frg2}
\end{eqnarray}
where $C(\Delta y,\Delta\phi)=\int_{|y|\leq \Delta y/2}dy
\int_{|\phi-\bar{\phi}_{\gamma}|
\leq \Delta\phi/2}d\phi f(y,\phi-\bar{\phi}_{\gamma})$ 
can be considered as an overall acceptance factor for finding 
the jet fragments in the given kinematic range.

We will approximate the fractional production cross section,
$r_a(E_T^{\gamma})$, of $a$-type jet associated with the direct 
photon, by the lowest order calculation,
\begin{eqnarray}
  r_a(E_T^{\gamma})&=&\frac{d\sigma^{\gamma}_a/dy_\gamma dE_T^{\gamma}}{
    d\sigma^{\gamma}/dy_\gamma dE_T^{\gamma}}\; ; \;\;
  \sigma^{\gamma} = \sum_a \sigma^{\gamma}_a\; ; \\
  \frac{d\sigma^{\gamma}_a}{dy_\gamma dE_T^{\gamma}} &=&
  \sum_{bc}\int_{x_{b{\rm min}}}^1 dx_b f_{b/p}(x_b) f_{c/p}(x_c)
  \frac{2}{\pi} \nonumber \\
  & \times & \frac{x_b x_c}{2x_b -x_Te^{y_\gamma}} 
  \frac{d\sigma}{d\hat{t}}(bc\rightarrow \gamma+a) ,
\end{eqnarray}
where $x_c=x_bx_Te^{-y_\gamma}/(2x_b-x_Te^{y_\gamma})$, 
$x_{b {\rm min}}=x_Te^{y_\gamma}/(2-x_Te^{-y_\gamma})$,
and $x_T=2E_T/\sqrt{s}$. The parton distributions in a proton,
$f_{a/p}(x)$, will be given by the MSRD$-\prime$ parametrization \cite{mrs}.
In our following calculations for $AA$ collisions, we will use
the impact-parameter averaged parton distributions per nucleon 
in a nucleus $(A,Z)$,
\begin{equation}
  f_{a/A}(x)=S_{a/A}(x)\left[\frac{Z}{A}f_{a/p}(x)
  +(1-\frac{Z}{A})f_{a/n}(x)\right] \; ,
\end{equation}
where $S_{a/A}(x)$ is the parton nuclear shadowing factor which 
we will take the HIJING parametrization \cite{hijing}.
We have explicitly taken into account the isospin of the nucleus
by considering the parton distributions of a neutron which are 
obtained from that of a proton by isospin symmetry.

\begin{figure}
\centerline{\psfig{figure=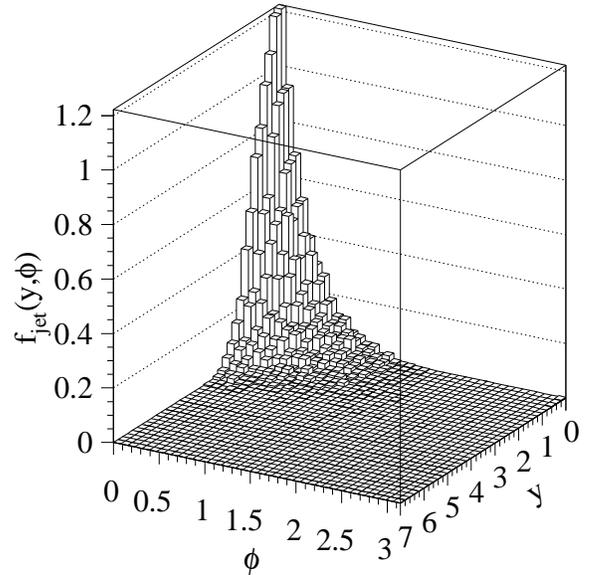,width=3in,height=3.0in}}

\caption{The normalized parton correlation from HIJING simulations
  in rapidity $y$ and azimuthal angle $\phi$ with respect to 
  the opposite direction of a tagged photon with $E_T^\gamma=10$ GeV 
  in $pp$ collisions at $\protect\sqrt{s}=200$ GeV}
\label{fig1}
\end{figure}

To simulate higher order effects, event generators such as 
PYTHIA \cite{pythia}, which was used in HIJING program \cite{hijing},
normally use parton shower model. In this model, one introduces a 
cut-off $\mu_0$ for the parton virtuality in the chain of parton shower,
thus avoiding both infrared and collinear singularities. In addition,
initial and final state radiations are treated separately and the
interference between them is also neglected.
Shown in Fig.~\ref{fig1}, is the parton correlation 
function in rapidity $y$ and azimuthal angle $\phi$
with respect to the opposite direction of a triggered photon
with $E^{\gamma}_T=10$ GeV in HIJING \cite{hijing} simulations
of $pp$ collisions at $\sqrt{s}=200$ GeV. In the simulations,
the final state radiations are switched off so that we
can study the effect of momentum imbalance due to initial
state radiations. We can see that most 
of the jet partons fall into the kinematic region,
$(|y|\leq 1, |\phi-\bar{\phi}_{\gamma}|\leq 0.5)$.
The jet profile, which is the convolution of the parton
correlation function and hadron intrinsic profile around
the parton axis, has a similar shape with a slightly larger 
width according to both our simulations and experimental
measurements in high-energy $p\bar{p}$ collisions \cite{ua1}.
For calculations throughout this paper, we will use $\Delta y=1$
and $\Delta \phi=2$. We find the acceptance factor defined
in Eq.~(\ref{eq:frg2}), $C(\Delta y,\Delta\phi)\approx 0.5$,
independent of both the colliding energy and the photon energy
$E_T^{\gamma}$, using HIJING \cite{hijing} Monte Carlo simulations.

\begin{figure}
\centerline{\psfig{figure=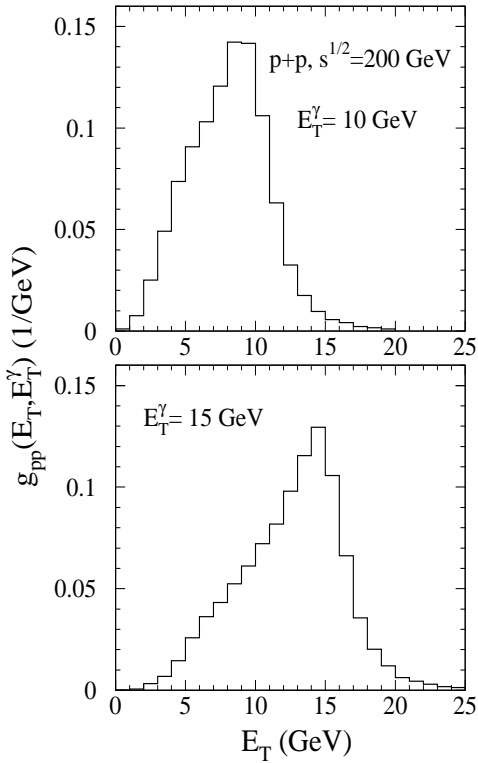,width=2.5in,height=4in}}

\caption{The $E_T$-smearing function for the photon-tagged parton jets
with $E_T^\gamma=10$, 15 GeV, from HIJING simulations of $pp$
collisions at $\protect\sqrt{s}=200$ GeV. The averaged value of
$E_T$ is $\langle E_T\rangle =8.08$, 12.57 GeV, respectively}
\label{fig2}
\end{figure}

Since a parton jet in the parton shower model is well defined,
one can then calculate the double differential cross section
for $\gamma-{\rm jet}$ events. Shown in Fig.~\ref{fig2} are the 
normalized $E_T$ distributions of the jet parton,
\begin{equation}
  \frac{1}{N_{\gamma-{\rm jet}}}
      \frac{dN_{\gamma-{\rm jet}}}{dE_T}\; , \label{eq:smear}
\end{equation}
with given $E_T^{\gamma}$ of the tagged direct photon, from 
HIJING simulations. As we can see that the transverse energy of 
the jet parton has a wide smearing around $E_T^{\gamma}$ due to the initial
state radiations associated with the hard processes. Because
of the rapidly decrease with $E_T$ of the direct photon production
cross section, the distribution is biased toward 
smaller $E_T$ than $E_T^{\gamma}$. The average $E_T$
is thus smaller than $E_T^{\gamma}$. In this paper, we
will use the jet parton distribution in Eq.~(\ref{eq:smear}) to
approximate the $E_T$-smearing function in Eq.~(\ref{eq:smear0}).

If one triggers a direct photon with a given  $E_T^{\gamma}$, one should
average over the $E_T$ smearing of the jet in the calculation
of particle distributions in the opposite direction of the 
tagged photon. Such a smearing is important especially for 
hadrons with $p_T$ comparable or larger than $E_T^{\gamma}$.

If we define the inclusive fragmentation function associated
with a direct photon in $pp$ collisions as,
\begin{eqnarray}
  D^{\gamma}_{pp}(z)&=&\sum_{a,h}r_a(E_T^{\gamma})
  \int dE_T g_{pp}(E_T,E_T^{\gamma}) \nonumber \\
  & \times & \frac{E_T^{\gamma}}{E_T}
  D_{h/a}^0(z\frac{E_T^{\gamma}}{E_T})\; , \label{eq:frg3}
\end{eqnarray}
with $z=p_T/E_T^{\gamma}$ the hadrons' momenta as fractions of 
the direct photon's transverse energy, we can rewrite the $p_T$ 
spectrum [Eq.~(\ref{eq:frg2})] of hadrons in the opposite direction of a 
tagged photon as
\begin{equation}
  \frac{dN_{pp}^{\gamma-h^\pm}}{dyd^2p_T}=
  \frac{D^{\gamma}_{pp}(p_T/E_T^{\gamma})}{p_T E_T^{\gamma}} 
  \frac{C(\Delta y,\Delta\phi)}{\Delta y\Delta\phi}\; . \label{eq:frg4}
\end{equation}

Considering parton energy loss in central $AA$ collisions,
we model the modified jet fragmentation functions as given by
Eq.~(\ref{eq:frg1}). Including the $E_T$ smearing
and averaging over the $\gamma$-jet production position
in the transverse direction, the inclusive fragmentation 
function of a photon-tagged jet in central $A+A$ collisions is,
\begin{eqnarray}
  D^{\gamma}_{AA}(z) &=&\int \frac{d^2r t^2_A(r)}{T_{AA}(0)}\sum_{a,h} 
  r_a(E_T^{\gamma})\int dE_T g_{AA}(E_T,E_T^{\gamma}) \nonumber \\
  &\times & \frac{E_T^{\gamma}}{E_T}
   D_{h/a}(z\frac{E_T^{\gamma}}{E_T},L) \; ,\label{eq:frg-aa}
\end{eqnarray}
where $T_{AA}(0)=\int d^2r t^2_A(r)$ is the overlap function of
$AA$ collisions at zero impact-parameter. The $E_T$-smearing function
$g_{AA}(E_T,E_T^{\gamma})$ in $AA$ collisions should be
different from that in $pp$ collisions due to initial multiple
parton scatterings. However, for the moment, we will regard them
as the same and postpone the discussion of the difference 
to the next section. We have assumed that direct
photon production rate is proportional to the number of binary 
nucleon-nucleon collisions. Neglecting expansion in the
transverse direction, the total distance a parton produced
at $(r,\phi)$ will travel in the transverse direction is
$L(r,\phi)=\sqrt{R_A^2-r^2(1-\cos^2\phi)}-r\cos\phi$.
Using the above inclusive fragmentation function in 
Eq.~(\ref{eq:frg4}), one can similarly calculate the $p_T$ 
spectrum of particles in the opposite direction of a tagged 
photon in $AA$ collisions. Setting the parton energy loss $dE/dx=0$,
the above equation should be reduced to the inclusive fragmentation 
function in $pp$ collisions in Eq.~(\ref{eq:frg3}) and the
corresponding $p_T$ spectrum should also become the same
as in $pp$ collisions.

To measure the modification of the inclusive fragmentation function
in experiments, one should first select events with a direct photon 
of energy $E_T^{\gamma}$.  Then one measures the particle spectrum 
in the kinematical region $(\Delta y, \Delta\phi)$ in the opposite 
direction of the tagged photon. After subtracting the background 
which is essentially the $p_T$ spectrum in ordinary events, one 
can use Eq.~(\ref{eq:frg4}) to extract the inclusive jet 
fragmentation function, $D^{\gamma}(z)$, from the resultant spectrum. 
One can then compare the extracted inclusive jet fragmentation 
function in central $AA$ collisions to that in $pp$ or 
peripheral $AA$ collisions to study the modification due to 
parton energy loss. Note that the centrality requirements for 
the signal ($\gamma+{\rm jet}$) and background (ordinary) events 
should be the same. The overall acceptance factor $C(\Delta y,\Delta\phi)$
in $AA$ collisions remains approximately the same as in $pp$
collisions with small but measurable corrections due to
the $k_T$ broadening of the leading parton as we will discuss later.

\begin{figure}
\centerline{\psfig{figure=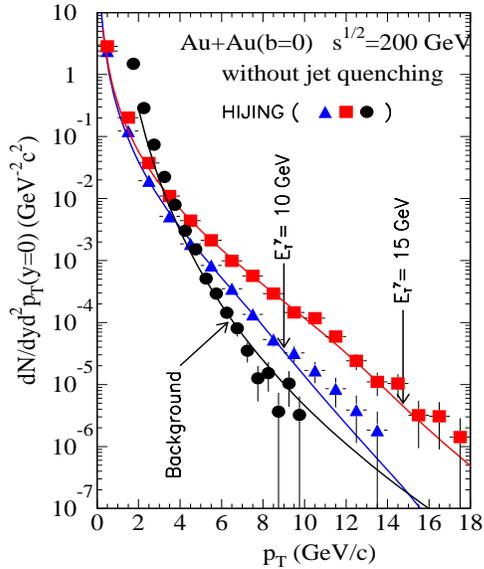,width=2.5in,height=3.0in}}

\caption{The differential $p_T$ spectrum of charged particles
  from the fragmentation of a photon-tagged jet with 
  $E_T^{\gamma}=$10, 15 GeV and the underlying 
  background in central $Au+Au$ collisions at
  $\protect\sqrt{s}=200$ GeV. The direct photon is restricted to 
  $|y|\leq \Delta y/2=0.5$. Charged particles are limited to the
  same rapidity range and in the opposite direction of the photon,
  $|\phi-\phi_{\gamma}-\pi|\leq \Delta\phi/2=1.0$. Solid lines
  are calculations using Eq.~(\protect\ref{eq:frg2}) and points are HIJING
  simulations of 20K events.}
\label{fig3}
\end{figure}

To demonstrate the feasibility of the above prescribed procedure,
we show in Fig.~\ref{fig3} the calculated $p_T$ distributions
of charged hadrons from the fragmentation of photon-tagged jets with 
$E_T^{\gamma}=10$, 15 GeV and the underlying background from
the rest of a central $Au+Au$ collisions at the RHIC energy.
We set $dE/dx=0$ so the effect of parton energy loss is not included yet.
The points are HIJING simulations of 20K events and solid lines
for jet fragmentation are numerical results from Eq.~(\ref{eq:frg4})
with the fragmentation functions given by the parametrization
of $e^+e^-$ data \cite{bkk}. The numerical result (solid line) 
for the background coming from jet fragmentation in ordinary
central events is obtained by the convolution of the jet cross
section and fragmentation functions \cite{wang2}. As we can see,
the spectra from jet fragmentation are significantly higher than
the background at moderately large transverse momenta. The background in $pp$
collisions is about 1200 (the number of binary nucleon-nucleon 
collisions) times smaller than in central $Au+Au$ collisions. 
One can therefore easily extract the inclusive fragmentation 
function from the experimental data without much statistical errors 
from the subtraction of the background. This conclusion remains valid 
even if one includes the parton energy loss in $AA$ collisions
because both the background and the particles from jet fragmentation
are suppressed by approximately the same amount due to jet 
quenching \cite{wang2,whs}. As one can expect from Fig.~\ref{fig3},
for direct photons with $E_T^\gamma < 6 $ GeV at the RHIC energy,
the hadron spectrum from the jet fragmentation is much smaller
than the background. In this case, one can no longer accurately 
extract the effective fragmentation function with finite number
of events.

\begin{figure}
\centerline{\psfig{figure=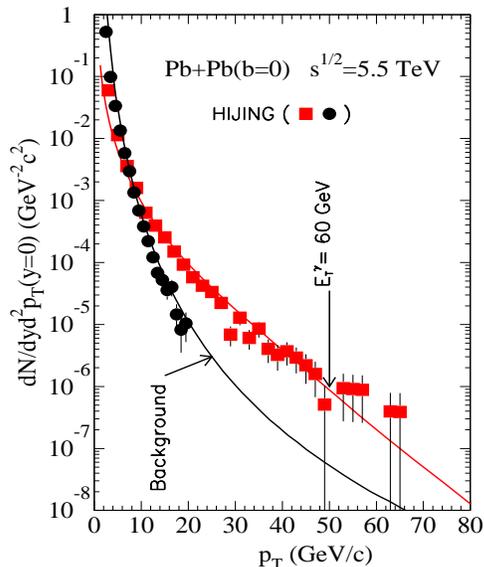,width=2.5in,height=3.0in}}

\caption{The same as Fig.~\protect\ref{fig3}, except for $E_T^\gamma=60$
  GeV at $\protect\sqrt{s}=5.5$ TeV.}
\label{fig4}
\end{figure}

Shown in Fig~\ref{fig4}, are similar calculations for central
$Pb+Pb$ collisions at the LHC energy with $E_T^\gamma =60$ GeV
for the tagged photons. It is clear that the overall background
is much larger than at the RHIC energy. Therefore, one needs to trigger
on large $E_T^\gamma$ photons. Our calculation shows that the
fragmented hadron spectrum from the photon-tagged jets with 
$E_T^\gamma=40$ GeV is roughly as large as the background at
the LHC energy, from which one can barely extract the effective
fragmentation function. For $E_T^\gamma <40$ GeV, the fragmentation
spectrum is too small to be extracted.

As a general criterion on the minimum value of
$E_T^\gamma$ in our prescribed procedure in central $AA$ collisions,
one should require
\begin{equation}
  E_T^\gamma>E_{T{\rm min}}^\gamma,\;\; 
  T_{AA}(0)\frac{d\sigma_{\rm jet}}{dydE_T}(E_{T{\rm min}}^\gamma)=1
  ( {\rm GeV}^{-1}).
\end{equation}

Since large $p_T$ hadrons in the background also come from
jet fragmentation, this is to ensure that $E_T^\gamma$ is large enough
such that the average number of jets with $E_T=E_T^\gamma$ in each 
central collisions is less than 1.
Only then, the inclusive fragmentation function of the photon-tagged
jet can be extracted with confidence after the subtraction of the
background. Shown in Table~\ref{tab0}, are values of $E_{T{\rm min}}^\gamma$
for different central $A+A$ collisions at $\sqrt{s}=200$ GeV.
This is consistent with what one can expect from Fig.~\ref{fig3}.

\begin{center}
\begin{table}
\begin{tabular}{|l|llll|}
$A$  & 80 & 120 & 160 & 200 \\ \hline
$E_{T{\rm min}}^\gamma$ (GeV) & 5.0 & 5.5 & 6.0 & 6.4 \\
\end{tabular}
\bigskip
\caption{The minimum transverse energy of the triggered photon,
  $E_{T{\rm min}}^\gamma$, required in order for the fragmentation
  function of photon-tagged jets to be reliably extracted in central
  $A+A$ collisions at $\protect\sqrt{s}=200$ GeV.}
\label{tab0}
\end{table}
\end{center}

\section{Effects of Jet $E_T$ Smearing}

\begin{figure}
\centerline{\psfig{figure=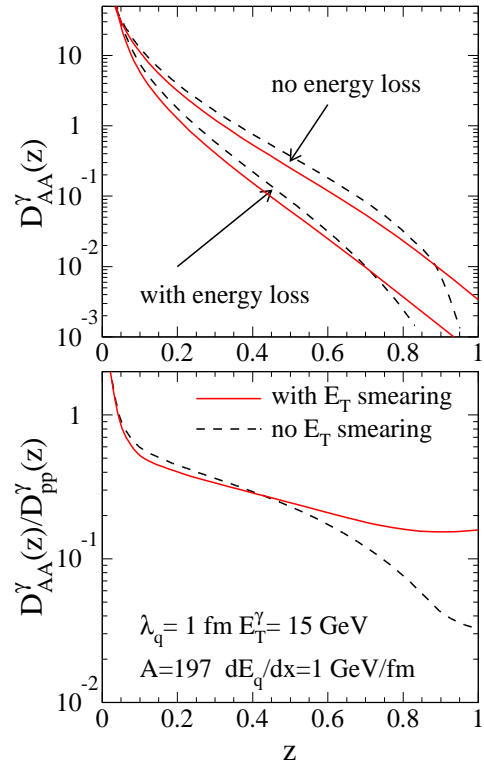,width=2.5in,height=4in}}

\caption{Upper panel:The inclusive fragmentation functions  
  with (solid lines) and without (dashed lines) $E_T$ smearing 
  for $dE_q/dx=1$ GeV/fm  and $dE_q/dx=0$.
  Lower panel: The ratios of the inclusive fragmentation 
  functions with and without energy loss. The mean-free-path 
  of a quark is assumed to be $\lambda_q=1$ fm and the 
  triggered photon has $E_T^{\gamma}=15$ GeV in central $Au+Au$
  collisions at $\protect\sqrt{s}=200$ GeV. }
\label{fig5}
\end{figure}

{}From Figs.~\ref{fig3} and \ref{fig4}, one also notices that 
there are significant number of particles with $p_T$ larger than 
$E_T^{\gamma}$, from fragmentation of the photon-tagged jets.
This is because of the $E_T$-smearing of
the jet caused by initial state radiations. To illustrate the 
effect of the $E_T$ smearing, we plot in Fig.~\ref{fig5} the 
inclusive fragmentation functions (upper panel) with (solid lines) 
and without (dashed lines) $E_T$ smearing both for $dE_q/dx=1$ GeV/fm  
and $dE_q/dx=0$. The lower panel shows the ratios of the inclusive 
fragmentation functions with and without energy loss. We assume 
that the mean-free-path of a quark is $\lambda_q=1$ fm and the 
triggered photon has $E_T^{\gamma}=15$ GeV in central $Au+Au$ 
collisions at $\sqrt{s}=200$ GeV. 
Notice that we now define $z$ as a hadron's fractional 
energy of the triggered photon. Because of the $E_T$-smearing
of the jet caused by initial state radiations, hadrons can
have $p_T$ larger than $E_T^{\gamma}$. Therefore, the effective
inclusive jet fragmentation function does not vanish at
$z=p_T/E_T^{\gamma} > 1$. As we can see, the effect of $E_T$-smearing
is only significant at large $z$. In particular at $z \simeq 1$,
the modified fragmentation function without $E_T$ smearing has 
contributions only from those partons which escape the system 
without a single inelastic scattering, thus is controlled only
by the mean-free-path \cite{whs}. However, after taking into
account of the $E_T$ smearing, one also has contributions
from the fragmentation of jets with $E_T$ larger than
$E_T^{\gamma}$ even if the jet has suffered energy loss. Therefore,
the modification of the inclusive fragmentation function in this
region of $z$ depends on both the mean-free-path and the parton 
energy loss.  Only at very large $z> 2$, the modification factor
becomes independent of the energy loss, depending only on the
mean-free-path. However, the production rate becomes also extremely
small. For small and intermediate values of $z$, both 
the inclusive fragmentation function and the modification
due to parton energy loss are not very sensitive to the $E_T$
smearing.

\begin{figure}
\centerline{\psfig{figure=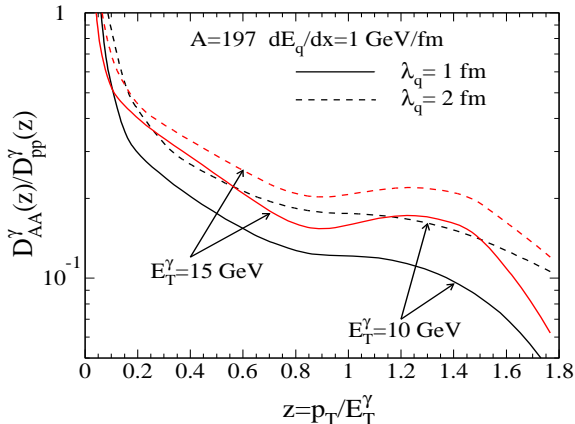,width=3in,height=2.25in}}

\caption{The modification factor of the photon-tagged inclusive jet
  fragmentation function in central $Au+Au$ collisions at 
  $\protect\sqrt{s}=200$ GeV for a fixed $dE_q/dx=1$ GeV/fm}
\label{fig6}
\end{figure}

To study the effect of jet $E_T$-smearing in detail, we show 
in Fig.~\ref{fig6} ratios of the inclusive fragmentation 
function in central $Au+Au$ collisions with energy 
loss $dE_q/dx=1$ GeV/fm over the one in $pp$ collisions 
without energy loss. We shall refer to this ratio as the {\em modification
factor}. The enhancement of soft particle production 
due to induced emissions is important only at very small fractional 
energy $z$. The fragmentation function is suppressed for large and 
intermediate $z$ due to parton energy loss. For fixed $dE/dx$ and $z$,
the suppression becomes less as $E_T^{\gamma}$ or the average 
$\langle E_T\rangle$ increases.  One can notice that there is
an interesting structure in the region of $z > 0.8$ which is
also a consequence of the jet $E_T$-smearing.  In this region,
contributions from fragmentation of jets with $E_T$ larger
than $E_T^{\gamma}$ dominates. Since the leading particles are
relatively less suppressed for larger $E_T$ with fixed $dE/dx$ and $z$,
the decrease of the modification factor will then saturate at
around $z\approx 1$ until the large $E_T$ tail of the $E_T$-smearing 
for the photon-tagged jet (as shown in Fig.~\ref{fig2})
becomes insignificant. Therefore, the structure of the modification
factor in the large $z$ region results from the competition between
the energy dependence of jet quenching and the falling-off of the
tail of the jet $E_T$ smearing. The structure becomes more
prominent for larger $E_T^{\gamma}$ and it should also depend
on the energy dependence of the parton energy loss.

Another advantage of studying jet quenching in photon-tagged
events is that the results are not sensitive to some of the
effects of initial state multiple interactions, {\it e.g.}, the
nuclear shadowing of parton distributions \cite{qn2} and
energy loss of the beam partons, which can affect the
direct photon production rate. However, there
is one exception, {\it i.e.}, the $E_T$ broadening from
multiple initial and final state scatterings with the beam partons. 
Such $E_T$ broadening, which also causes the so-called
Cronin effect \cite{cronin1} in $pA$ collisions, should also
increase the $E_T$ smearing of the photon-tagged jets in
both $pA$ and $AA$ collisions.  This $E_T$ broadening has also been 
seen in dijet events of fixed target experiments \cite{cronin3}.
Even though the Cronin effect
for inclusive cross sections decreases with the colliding energy 
as indicated by current experiments \cite{cronin2} and  theoretical 
estimates also predict it to be small \cite{report} at the RHIC collider
energy and beyond, the small $E_T$ broadening in the photon-tagged 
events should still have finite effects and one would 
like to have a handle on it experimentally. 
 
One can directly measure the $E_T$ distributions of the photon-tagged
jets in $pp$ and $pA$ collisions and study the possible change,
using the conventional calorimetrical study of jets.
However, due to small but finite background in $pp$ and $pA$
collisions, it is difficult to measure the calorimetric energy
of jets with an accuracy of less than 1 GeV. 
Here we propose to study the $E_T$ broadening indirectly
by measuring the modification factor for the photon-tagged jets
in $pA$ collisions, since the inclusive fragmentation function 
is very sensitive to the $E_T$ smearing as we have demonstrated
and the final jet partons in the central rapidity region
do not experience transverse energy loss 
in $pA$ collisions. In particular at $z=p_T/E_T^\gamma \sim 1$,
only those jets with $E_T>E_T^\gamma$ contributes. The spectrum 
in this region is extremely sensitive to the $E_T$ broadening.
Therefore, one should be able to measure even very small $E_T$ broadening
via the modification factor in $pA$ collisions.

Let us assume that the transverse momentum kick from initial
and final state scatterings with the beam partons has a Gaussian
distribution with a width $\Delta_{pA}$, one can then calculate
the effective jet fragmentation function similarly 
to Eq.~(\ref{eq:frg3}) but with a modified $E_T$-smearing function
$g_{pA}(E_T,E_T^{\gamma})$,
\begin{eqnarray}
  D^{\gamma}_{pA}(z)&=&\sum_{a,h}r_a(E_T^{\gamma})
  \int dE_T g_{pA}(E_T,E_T^{\gamma}) \nonumber \\
  & \times & \frac{E_T^{\gamma}}{E_T}
  D_{h/a}^0(z\frac{E_T^{\gamma}}{E_T})\; . \label{eq:frg-pa}
\end{eqnarray}
The modified $E_T$-smearing function can be obtained as the
convolution of the $E_T$-smearing function in $pp$ collisions
with a Gaussian distribution,
\begin{eqnarray}
g_{pA}(E_T,E_T^\gamma)&=&\int dE^a_T \int \frac{d^2p_T}{\pi\Delta_{pA}^2}
  e^{-p_T^2/\Delta_{pA}^2}g_{pp}(E^a_T,E_T^\gamma) \nonumber \\
  &\times& \delta\left(E_T-\sqrt{{E^a_T}^2+p_T^2+2p_T E^a_T\cos\phi}\right)
  \nonumber \\
  &=&\int_0^{\pi}\frac{d\phi}{\pi}\int_0^{E^2_T} 
  \frac{dp^2_T}{\Delta_{pA}^2} g_{pp}(E^a_T,E_T^\gamma)
  \nonumber \\ &\times&e^{-p_T^2/\Delta_{pA}^2}
  \frac{E_T}{\sqrt{{E_T}^2-p_T^2(1-\cos^2\phi)}}, \label{eq:sm}
\end{eqnarray}
where $E^a_T=\sqrt{{E_T}^2-p_T^2(1-\cos^2\phi)}-p_T\cos\phi$.
For not very large values of $\Delta_{pA}$ relative to $E_T^\gamma$,
the peak of the modified smearing function $g_{pA}(E_T,E_T^\gamma)$
is simply shifted to larger values of $E_T$ as compared to
$g_{pp}(E_T,E_T^\gamma)$ . One can characterize the $E_T$-shift by
\begin{equation}
  \Delta E_T=\int dE_T E_T[g_{pA}(E_T,E_T^\gamma)
  -g_{pp}(E_T,E_T^\gamma)] .
\end{equation}

\begin{figure}
\centerline{\psfig{figure=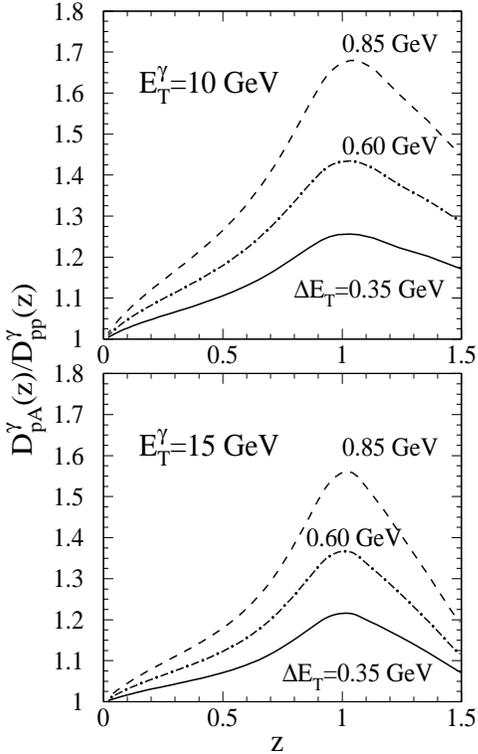,width=2.5in,height=4in}}

\caption{The modification factor for the photon-tagged jet fragmentation
  function in $pA$ collisions with $E_T^\gamma=10$ and 15 GeV
  at $\protect\sqrt{s}=200$ GeV, for different values of $\Delta E_T$
  due to $E_T$ broadening.}
\label{fig7}
\end{figure}

To demonstrate the sensitivity of the effective fragmentation
function on the $E_T$ broadening due to multiple
parton scatterings, we show in Fig~\ref{fig7} the modification
factor $D^\gamma_{pA}(z)/D^\gamma_{pp}(z)$ in $pA$ collisions
for three values of $\Delta E_T$ with $E_T^\gamma=10$ and 15 GeV,
respectively, at the RHIC energy. It is clear that the modification
factor is sensitive to the additional $E_T$ smearing even for 
very small values of $\Delta E_T$. The shape of the modification
factor simply reflects the fact that the smearing function
is most modified around the peak $E_T\approx E_T^\gamma$.
Comparison between the calculation for $E_T^\gamma=10$ and 15 GeV
for fixed values of $\Delta E_T$ shows that the relative effect of 
multiple scatterings decreases with increasing $E_T^\gamma$.

\begin{figure}
\centerline{\psfig{figure=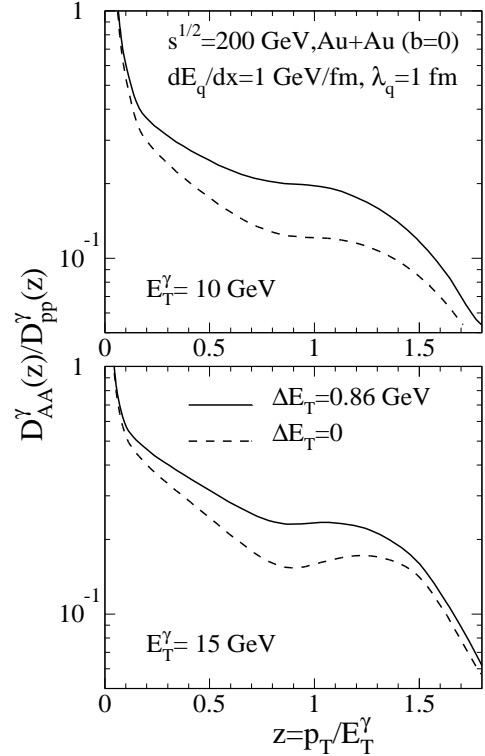,width=2.5in,height=4in}}

\caption{The modification factor for the inclusive fragmentation
  function of photon-tagged jets with (solid) and without (dashed)
  $E_T$ broadening due to initial parton scatterings, 
  in central $Au+Au$ collisions
  at $\protect\sqrt{s}=200$ GeV. The parton energy loss is fixed at
  $dE_q/dx=1$ GeV/fm and the mean-free-path $\lambda_q=1$ fm.}
\label{fig8}
\end{figure}

The $E_T$-smearing function for $AA$ collisions can be modeled
the same way as in Eq.~(\ref{eq:sm}) for $pA$ collisions, except
that $\Delta_{pA}$ is replaced by $\Delta_{AA}$. According to
the classical random-walk approximation \cite{lpm2,gv1}, 
$\Delta_{pA}^2$ should be proportional to $A^{1/3}$, 
or the average number of proton-nucleon subcollisions. 
In such an approximation, $\Delta_{AA}^2=2\Delta_{pA}^2$.
Using this modified $E_T$-smearing function in Eq.~(\ref{eq:frg-aa}),
we can calculate the modified effective fragmentation function
for the photon-tagged jets in $AA$ collisions, including both the
effect of parton energy loss through the dense medium and
the additional $E_T$-smearing due to initial and final state
scatterings with the beam partons. Shown in Fig.~\ref{fig8} are
the calculated modification factors with (solid line) and without
(dashed line) $E_T$ broadening. It is clear that the
$E_T$ broadening has significant effect on the final modification
factor in $AA$ collisions. One therefore has to study $pp$, $pA$ and $AA$ 
collisions systematically and subtract the effect caused 
by the $E_T$ broadening due to initial multiple scatterings to 
obtain the modification factor only due to parton energy loss 
in $AA$ collisions.
In the following discussions, we assume that such effect has already
been subtracted out and we only concentrate on the effect of 
parton energy loss.

\section{Extracting Parton Energy Loss}

Given the modification of the inclusive fragmentation
function of photon-tagged jets, one in principle should
be able to extract the parton energy loss and the parton
mean-free-path in our phenomenological model.
The optimal case is when the average total energy loss
is significant as compared to the initial jet energy,
and yet the $p_T$ spectrum from jet fragmentation is
still much larger than the underlying background.
However, because of the complication of the initial
state radiations, one still cannot determine precisely the 
energy of the photon-tagged jet in each central $AA$ event.
Therefore, the parton energy loss and mean-free-path cannot
be determined independently in a tangible way.
As compared to our earlier results \cite{whs} where
we did not take into account of the $E_T$ smearing of the
photon-tagged jets, the modification of the averaged fragmentation
function due to energy loss is quite sensitive to
the value of the mean-free-path for $dE_q/dx=$ 1 GeV as shown
in Fig.~\ref{fig6}. 

\begin{figure}
\centerline{\psfig{figure=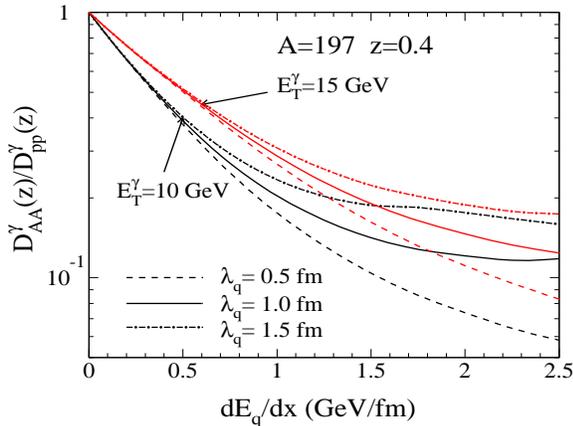,width=3in,height=2.25in}}

\caption{The modification factors for the inclusive fragmentation
  function of photon-tagged jets at given $z=p_T/E_T^\gamma=0.4$
  as functions of parton energy loss $dE_q/dx$ with different 
  values of the mean-free-path
  $\lambda_q$, in central $Au+Au$ collisions at $\protect\sqrt{s}=200$ GeV.}
\label{fig9}
\end{figure}

To study the sensitivity of the modification 
to the energy loss, we plot in Fig.~\ref{fig9} the modification
factor at a fixed value of $z=0.4$ as functions of
$dE_q/dx$.  For small values of $dE_q/dx$, the suppression factor is more
or less independent of the mean-free-path. This is referred to as
the ``soft emission'' scenario in Ref.~\cite{whs} where the suppression is 
dominated by the leading parton with an average total energy
loss $\langle \Delta E_T^a\rangle=\langle n_a \rangle\epsilon_a
=\langle L\rangle dE_a/dx$. The suppression factor 
should scale with $dE_a/dx$, depending very weakly on 
the mean-free-path. Assuming an exponential form of the fragmentation
function $\sim e^{-cz}$ for $z=0.2\sim 0.8$, one can show that 
the suppression factor has a form 
$(1-\Delta E_T^a/E_T^\gamma)\exp(-cz\Delta E_T^a/E_T^\gamma)$.
For large values of $dE_q/dx\geq 1$ GeV/fm, the ratio is sensitive
to the mean-free-path. However, as one can see from Fig.~\ref{fig9}, the
suppression factor flattens out as $dE_q/dx$ increases, especially for
large values of the mean-free-path $\lambda_a$.  This can be understood
as the ``hard emission'' scenario in which the parton enegy loss per
emission $\epsilon_a=\lambda_a dE_a/dx$ is large.
In this scenario, $z_n$ after $n$ times emission
becomes so small, the $n\neq 0$ contribution is completely  suppressed
in the modified fragmentation function in Eq.~(\ref{eq:frg1}). 
The only contribution is from $n=0$ term, {\it i.e.}, 
from the partons which escape the system without any induced 
radiation. In this case, the suppression factor is controlled by 
$\exp(-L/\lambda_a)$, a factor independent of $dE_a/dx$ and $E_T^\gamma$.
One thus needs to measure the suppression factor at smaller values 
of $z$ or a global fit to determine both the energy loss $dE_a/dx$ 
and the mean-free-path from the experimental data.

Recent theoretical studies \cite{lpm2} of parton energy loss 
in a dense medium of a finite size $L$ indicate that
the energy loss per unit distance $dE/dx$ could be proportional
to the total distance that the parton has traveled since
it is produced,
\begin{eqnarray}
  \frac{dE_a}{dx}&=&\frac{C_a\alpha_s}{8} 
  \frac{\mu^2}{\lambda} L \ln\frac{ L}{\lambda} \nonumber \\
  &=&\frac{C_a\alpha_s}{8} \Delta k_T^2 \; , \label{eq:ptL}
\end{eqnarray}
which was also shown to be proportional to the average transverse
momentum broadening squared, $\Delta k_T^2$, with respect to the
direction of the initial parton momentum, where $\mu$ is the Debye
mass of the medium and $\lambda$ the mean-free-path of the parton,
$C_a=$ 4/3 for quark and 3 for a gluon.
The $k_T$ broadening results from multiple scatterings which also 
induce the radiative energy loss for the propagating parton. 

\begin{figure}
\centerline{\psfig{figure=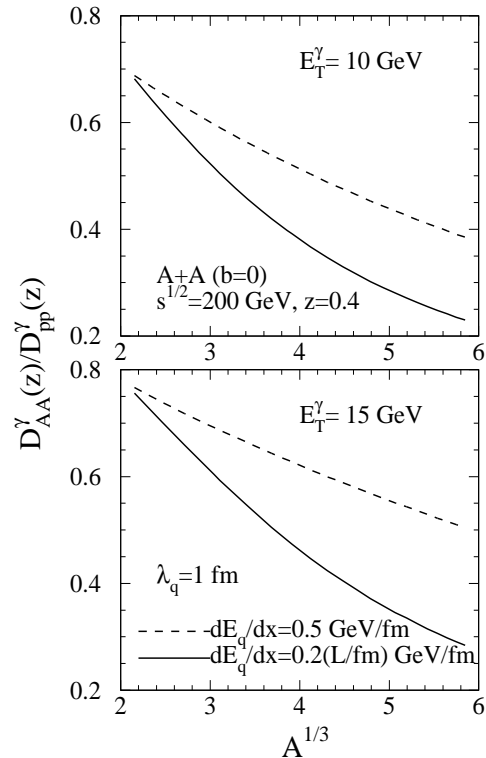,width=2.5in,height=4in}}

\caption{The modification factors for the inclusive fragmentation
  function of photon-tagged jets at given $z=p_T/E_T^\gamma=0.4$
  as functions of $A^{1/3}$ in central $A+A$ collisions at 
  $\protect\sqrt{s}=200$ GeV. The solid lines are for a distance-dependent
  parton energy loss $dE_q/dx$, while the dashed lines are for
  a constant $dE_q/dx$.}
\label{fig10}
\end{figure}

One way to test this experimentally is to study the modification 
factor at any given $z$ value for different nucleus-nucleus collisions 
or for different centralities (impact parameters). Shown in Fig.~\ref{fig10},
are the modification factors for the inclusive fragmentation function
at $z=0.4$ as functions of $A^{1/3}$. We assume that the radius of the
cylindrical system is $R_A=1.2 A^{1/3}$ fm. In one case (dashed lines),
we assume a constant energy loss $dE/dx$=0.5 GeV/fm. The 
modification factor decreases almost linearly with $A^{1/3}$.
In another case (solid lines), we assume $dE_q/dx=0.2 (L/{\rm fm})$ GeV/fm.
The average transverse distance a parton travels in a cylindrical system
with transverse size $R_A$ is $\langle L\rangle =0.905 R_A$.
We choose the coefficient in $dE_q/dx$ such that its average value
roughly equals 
to 0.5 GeV/fm for $A=20$. To implement such an energy loss in
our model, we assume the energy loss per scattering, for a parton 
traveling a total distance $L$, to be 
$\epsilon_a=\lambda_a 0.2 (L/{\rm fm})$ GeV.
As we can see, the suppression factor for a 
distance-dependent $dE/dx$ decreases faster than
the one with constant $dE/dx$. Unfortunately, we have
not found a unique way to extract the average total
energy loss so that one could show that it is proportional
to $A^{2/3}$ for the distance-dependent $dE/dx$. One
possible procedure to determine the $A$ dependence
of the energy loss is to first determine $dE/dx$ and $\lambda$
using a global fit to the measured modification factor for
each type of central $AA$ collisions and then find the
$A$ dependence of the extracted $dE/dx$. However, such a procedure
and our model depend on the assumption of the size of
the dense medium produced in $AA$ collisions.

\section{$k_T$ Broadening and Jet Profile}

In our discussions so far, we have assumed that the jet profile
in the opposite direction of the tagged photon remains the
same in $AA$ as in $pp$ collisions, since we used the same acceptance
factor $C(\Delta y, \Delta\phi)$. Such an acceptance factor
is determined by the effective jet profile in the opposite
direction of the tagged photon. One can imagine that there should 
be two sources of corrections. One is due to the initial and final
state multiple parton scatterings with the colliding nucleons. 
As we have discussed,
such multiple scatterings can cause the broadening of the
jet $E_T$ smearing. They shall also increase the acoplanarity
of the jet with respect to the tagged photon. One can study
this effect directly via the effective jet profile in
$pA \rightarrow \gamma +{\rm jet} + X$  processes as in dijet
events \cite{cronin3}. Let
us assume that such increased acoplanarity can be measured and corrected. 
The second correction to the effective jet profile comes from multiple
scatterings suffered by the leading parton while it propagates
inside the dense medium. These multiple scatterings induce
radiative energy loss and in the meantime also cause the
$k_T$ broadening of the final parton with respect to its
original transverse direction, giving rise to
an additional acoplanarity . Such a change to the jet
profile could affect the acceptance factor, which will be
an overall factor to the measured jet fragmentation function
if we assume the jet profile to be the same for particles
with different fractional energies.

Since we only consider jets in the central rapidity region ($y=0$),
we assume that the final multiple scatterings will only change
the jet profile in the azimuthal direction. We define the
jet profile function as $f(\phi)=dE_T/d\phi$. If the initial
effective jet profile is $f_0(\phi)$ and the $k_T$ broadening 
distribution is given by a Gaussian form \cite{lpm2}, the final 
effective jet profile function is then,
\begin{equation}
  f(\phi)=\int_0^{\infty} dk_T^2 \frac{1}{\Delta k_T^2}
  e^{-k_T^2/\Delta k_T^2} f_0(\phi-\phi_{jet})\; , \label{eq:profile}
\end{equation}
where $\sin\phi_{jet}=k_T/E_T$ and $\Delta k_T^2$ is the
average $k_T$ broadening squared.

\begin{figure}
\centerline{\psfig{figure=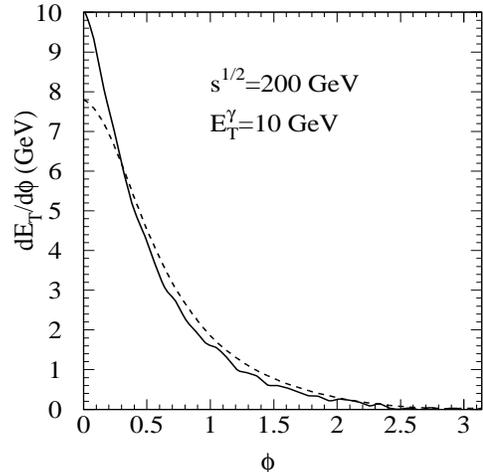,width=2.5in,height=2.5in}}

\caption{Jet profile $dE_T/d\phi$ (within $|y|<0.5$) with respect
  to the opposite direction of the tagged photon. The solid line
  is the original profile in $pp$ collisions from HIJING simulations
  while the dashed line
  is the modified profile function with $\Delta k_T^2=4$ GeV$^2/c^2$.}
\label{fig11}
\end{figure}

To demonstrate the effect of the $k_T$ broadening due to
final multiple scatterings, we plot in Fig.~\ref{fig11}
(solid line) the azimuthal angle distribution of $E_T$ (within $|y|<0.5$) 
with respect to the opposite direction of the tagged photon
with $E_T^{\gamma}=10$ GeV. We have subtracted the background 
so that $dE_T/d\phi=0$ at $\phi=\pi$. The profile distribution 
includes both the intrinsic distribution from jet fragmentation 
and the effect of initial state radiations. The acceptance factor 
is simply the fractional area within $|\phi|<\Delta\phi/2$ region.
The $k_T$ broadening of jets due to multiple scatterings
will broaden the profile function. Shown as the dashed line
is the profile function from Eq.~(\ref{eq:profile}) for 
$\Delta k_T^2=4$ (GeV/$c$)$^2$. It is clear that with a modest
value of the $k_T$ broadening, the acceptance factor only
changes by a few percents.

Since the change of the effective jet profile function 
is related to the average $k_T$ broadening, one can combine the measurement
with the measured energy loss to verify the relationship between
$dE/dx$ and $\Delta k_T^2$ as in Eq.~(\ref{eq:ptL}).

\section{Jet Quenching in Deep Inelastic Lepton-Nucleus Scatterings}

Even though we have so far applied the parton energy loss in 
Eq.~(\ref{eq:ptL}) to a fast parton inside a dense matter,
the generic form and its derivation is also valid for a
parton propagating inside a cold nuclear matter or hot hadronic
medium. The properties of the medium are manifested in the total
transverse momentum kick $\Delta k_T^2$. For a hot QGP, $\Delta k_T^2$
directly reflects the temperature, while for a cold nuclear matter
it is related to the gluon density inside a nucleus \cite{lpm2}.
If there is a dramatic difference between the transverse momentum
broadening or the parton energy loss in QGP and 
a cold nuclear matter, then the measurement of parton energy loss 
in high-energy heavy-ion collisions can be used as a possible probe
of QGP formation. It is thus also important to measure the
parton energy loss  in a cold nuclear matter.

As we have mentioned in the Introduction, initial state interactions
with beam nucleons prior to a hard process can also cause the
participating partons to lose energy, thus affecting the final
cross section. Among many hard processes, such as Drell-Yan lepton pair
and heavy quarkonium production at large $x_F$ in $pA$ 
collisions \cite{gm}, the simplest processes where parton
energy loss in cold nuclear matter can be directly measured
are probably deeply inelastic lepton-nucleus scatterings.
In such processes, one can relate the suppression of the leading
hadrons to the attenuation of the quark jet inside the
nuclear matter. There are many earlier studies of this problem
in the literature \cite{NAG,niko,bialas,gp_eA}. In this paper,
we would like to revisit this problem within our framework
of modified fragmentation functions.

In deeply inelastic $\ell A$ collisions, a quark or anti-quark 
is knocked out of a nucleon by the virtual photon which carries 
energy $\nu$ and virtuality $Q$. In the rest frame of the nucleus, 
the photon's energy is transferred to the quark which then will propagate
through the rest of the nucleus. If the hadronization time
of the quark in the order of $2\nu/\Lambda^2_{\rm QCD}$ is much larger
than the nuclear size, most of the leading hadrons from the jet 
fragmentation are formed outside of the nucleus. We can then
attribute the attenuation of the leading hadrons from the quark
fragmentation to the energy loss of the propagating quark 
inside the nuclear matter. Let us assume that the
longitudinal position of the nucleon from which the quark is knocked out
is $x_{\parallel}$ in the direction of the virtual photon.
Using Eq.~(\ref{eq:frg1}) for the modified fragmentation functions
due to parton energy loss and averaging over the longitudinal and
transverse position of the interaction point inside the nucleus, we can
obtain the effective quark fragmentation functions in deeply
inelastic lepton-nucleus collisions,
\begin{equation}
  D^{\ell A}_{h/a}(z,Q^2)=\frac{3}{4}\int_0^{R_A^2}\frac{dx^2_\perp}{R_A^2}
  \int_{-L_A(x_\perp)}^{L_A(x_\perp)}
  \frac{dx_\parallel}{R_A} D_{h/a}(z,\Delta L,Q^2)
\end{equation}
where $L_A(x_\perp)=\sqrt{R_A^2-x^2_\perp}$, 
$\Delta L=-x_\parallel +L_A(x_\perp)$, and a hard-sphere
nuclear distribution is used.

In a parton model, the nuclear structure function is defined as
\begin{eqnarray}
  F_2^{\ell A}(x,Q^2)&=&Z F_2^{\ell p/A}(x,Q^2)
  + (A-Z) F_2^{\ell n/A}(x,Q^2)\;\;, \nonumber \\
  \frac{1}{x}F_2^{\ell p/A}(x,Q^2)&=&\frac{4}{9}[2V_{N/A}(x,Q^2)
  +2\bar{u}_{N/A}(x,Q^2)] \nonumber \\
  &&\hspace{-1.0in}+\frac{1}{9}[V_{N/A}(x,Q^2)+2\bar{d}_{N/A}(x,Q^2)
  +2\bar{s}_{N/A}(x,Q^2)]\;\;,\nonumber \\
  \frac{1}{x}F_2^{\ell n/A}(x,Q^2)&=&\frac{4}{9}[V_{N/A}(x,Q^2)
  +2\bar{u}_{N/A}(x,Q^2)]  \nonumber \\
  &&\hspace{-1.0in}+\frac{1}{9}[2V_{N/A}(x,Q^2)+2\bar{d}_{N/A}(x,Q^2)
  +2\bar{s}_{N/A}(x,Q^2)]\;\;,
\end{eqnarray}
where $V_{N/A}(x,Q^2)$ (normalized to 1) is the effective 
valence quark distribution, 
$\bar{q}_{N/A}(x,Q^2)$'s are the effective sea quark distributions 
per nucleon inside a nucleus and $x=Q^2/2m_N\nu$.
Here we neglect the isospin asymmetry in the sea quark distributions,
{\it i.e.}, $\bar{u}_{N/A}(x,Q^2)=\bar{d}_{N/A}(x,Q^2)$. 
Because of nuclear effects such as shadowing,
the effective parton distributions per nucleon inside 
a nucleus are different from that inside a nucleon in the vacuum.
There are many different mechanisms for the nuclear modification
of the parton distributions and they could be different for
valence and sea quark distributions \cite{eskola}. In this paper,
we assume that the nuclear modification factors for quark distributions
are the same and can be given by the ratio of structure functions,
\begin{eqnarray}
  R_{A/D}(x,Q^2)&=&\frac{V_{N/A}(x,Q^2)}{V_{N/D}(x,Q^2)}
    =\frac{\bar{q}_{N/A}(x,Q^2)}{\bar{q}_{N/D}(x,Q^2)} \nonumber \\
    &=&\frac{2F_2^{\ell A}(x,Q^2)}{AF_2^{\ell D}(x,Q^2)}
\end{eqnarray}

\begin{figure}
\centerline{\psfig{figure=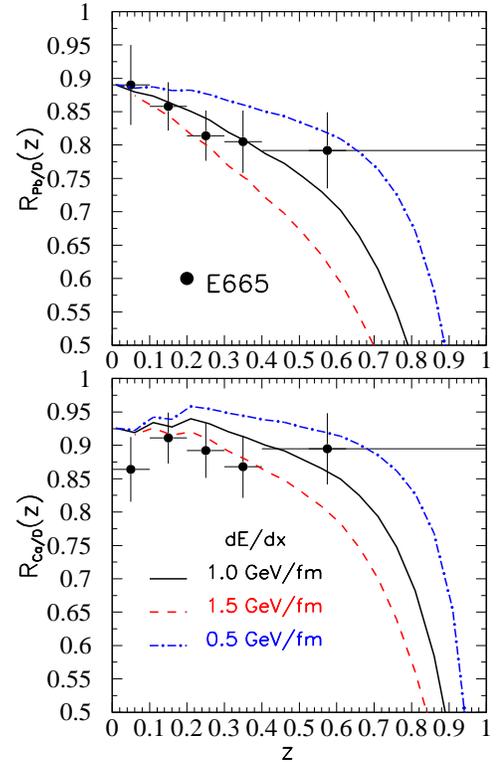,width=2.5in,height=4in}}

\caption{The suppression factor of charged hadron production
  in the jet fragmentation region in deeply inelastic $\ell A$
  collisions as a function of $z=E_h/\nu$. The lines are
  calculations using modified fragmentation functions with
  parton energy loss $dE_q/dx$. The data are from 
  Ref.~\protect\cite{e665}.}
\label{fig12}
\end{figure}

With parton fragmentation model, one can also define the 
semi-inclusive structure function associated with production 
of hadrons with momentum $z\nu$ in the direction of the virtual photon,
\begin{eqnarray}
  F_2^{\ell A\rightarrow h^\pm}(x,z,Q^2)&=&
  Z F_2^{\ell p/A\rightarrow h^\pm}(x,z,Q^2)\nonumber \\
  &+& (A-Z) F_2^{\ell n/A\rightarrow h^\pm}(x,z,Q^2)\;\;, \nonumber \\
  \frac{1}{x}F_2^{\ell p/A\rightarrow h^\pm}(x,z,Q^2)&=&
  \frac{4}{9}[2V_{N/A}(x,Q^2)+2\bar{u}_{N/A}(x,Q^2)]
  [D^{\ell A}_{\pi/V}(z,Q^2)+D^{\ell A}_{K/V}(z,Q^2)] \nonumber \\
  &+&\frac{1}{9}[V_{N/A}(x,Q^2)+2\bar{d}_{N/A}(x,Q^2)]
  [D^{\ell A}_{\pi/V}(z,Q^2)+D^{\ell A}_{K/S}(z,Q^2)] \nonumber \\
  &+&\frac{1}{9}2\bar{s}_{N/A}(x,Q^2)
  [D^{\ell A}_{\pi/S}(z,Q^2)+D^{\ell A}_{K/V}(z,Q^2)]\;\;,\nonumber \\
  \frac{1}{x}F_2^{\ell n/A\rightarrow h^\pm}(x,z,Q^2)&=&
  \frac{4}{9}[V_{N/A}(x,Q^2)+2\bar{u}_{N/A}(x,Q^2)]
  [D^{\ell A}_{\pi/V}(z,Q^2)+D^{\ell A}_{K/V}(z,Q^2)] \nonumber \\
  &+&\frac{1}{9}[2V_{N/A}(x,Q^2)+2\bar{d}_{N/A}(x,Q^2)]
  [D^{\ell A}_{\pi/V}(z,Q^2)+D^{\ell A}_{K/S}(z,Q^2)] \nonumber \\
  &+&\frac{1}{9}2\bar{s}_{N/A}(x,Q^2)
  [D^{\ell A}_{\pi/S}(z,Q^2)+D^{\ell A}_{K/V}(z,Q^2)]\;\;,\nonumber \\  
\end{eqnarray}
In principle, one should also take into account hadron
production from the fragmentation of the nuclear remnants. However,
one can neglect them for relatively large values of $z$.
In the above equation, we have used the following definitions
for the quark fragmentation functions,
\begin{eqnarray}
  D_{\pi/V}&\equiv&D_{\pi/u}=D_{\pi/\bar{u}}=D_{\pi/d}=D_{\pi/\bar{d}},
  \nonumber \\
  D_{\pi/S}&\equiv&D_{\pi/s}=D_{\pi/\bar{s}},\nonumber \\
  D_{K/V}&\equiv&D_{K/u}=D_{K/\bar{u}}=D_{K/s}=D_{K/\bar{s}}, 
  \nonumber \\
  D_{K/S}&\equiv&D_{K/d}=D_{K/\bar{d}}.
\end{eqnarray}

With the above model assumptions, one can study the quark 
energy loss by measuring the modification of
the jet fragmentation functions via the ratio of the above semi-inclusive 
structure functions for different nucleus ($A$ and $D$-deuterium) targets,
\begin{equation}
  R_{A/D}(z)=\frac{2F_2^{\ell A\rightarrow h^\pm}(x,z,Q^2)}
  {AF_2^{\ell D\rightarrow h^\pm}(x,z,Q^2)}
\end{equation}

Shown in Fig~\ref{fig12}, is the calculation of the above ratio
within our model of the modified fragmentation functions, together
with experimental data from E665 \cite{e665}. In the experiments,
the averaged values of $Q^2$, $\nu$ and $x$ are different for different
values of $z$. We have taken into account such kinematic effects
in our calculation, especially the nuclear modification of the
quark distributions. In the experimental measurements, small
values of $z$ are correlated to small values of $x$, where nuclear
shadowing of parton distributions is important. This is why
the ratio $R_{A/D}(z)$ is smaller than 1 even at small values of $z$.
Within the errors, the data are consistent with our calculation 
with parton energy loss of $dE/dx=0.5$--1 GeV/fm. It is clear that
in order to pin down the quark energy loss inside the nuclear matter,
one still needs more accurate measurements in deeply inelastic
$\ell A$ collisions. Because of the finite total parton energy
loss possible inside a finite nucleus, events with small values
of $\nu$ (energy carried by the struck quark) are more desirable. 

\vspace{2.5in}
\mbox{}\\

\section{Experimental Feasibilities}

To have an estimate of the experimental feasibility of our
proposed $\gamma+{\rm jet}$ measurement, we list in Table~\ref{tab1}
the number of $\gamma+{\rm jet}$ events per year per unit rapidity and
unit (GeV) $E_T$ at the RHIC collider energy. We assume a central $Au+Au$
cross section of 125 mb with impact-parameters $b<2$ fm.
We have taken a luminosity of ${\cal L}=2\times 10^{26}$
cm$^{-2}$s$^{-1}$ with 100 operation days per year.
The $\gamma+{\rm jet}$ cross sections are taken from the
compilation by the Hard Probes (HP) Collaboration \cite{hp1}
As we can see, although the number of direct photons with
$E_T^\gamma=7$ GeV is large enough,
the rate for $E_T^{\gamma}=15$, 20 GeV is still too
small to give any statistically significant measurement of
the fragmentation function and its modification in $AA$
collisions. If one can increase the luminosity by a factor
of 10, the numbers of events for both $E_T^{\gamma}=$10 and 15 GeV
are significant enough for a reasonable determination of
the fragmentation function of the photon-tagged jets.

\begin{figure}
\centerline{\psfig{figure=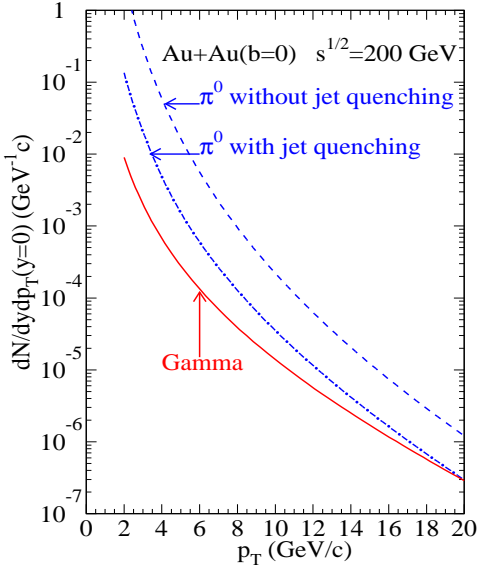,width=2.5in,height=3.0in}}

\caption{The spectrum of direct photon production (solid) as
  compared to $\pi^0$ spectrum with (dot-dashed) and without (dashed)
  parton energy loss ($dE_q/dx=1$ GeV/fm, $\lambda_q=1$ fm)
  in central $Au+Au$ collisions at $\protect\sqrt{s}=200$ GeV.}
\label{fig13}
\end{figure}

\begin{figure}
\centerline{\psfig{figure=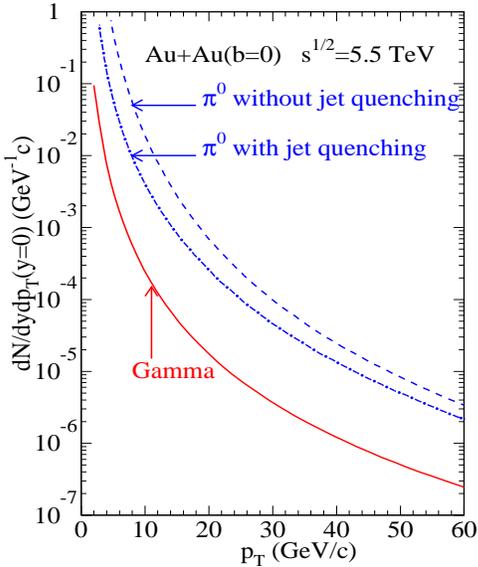,width=2.5in,height=3.0in}}

\caption{The same as Fig.~\protect\ref{fig13}, except at 
  $\protect\sqrt{s}=5.5$ TeV.}
\label{fig14}
\end{figure}

Given enough number of events, one still
has to overcome the large background of $\pi^0$'s to identify
the direct photons. Plotted in Figs.~\ref{fig13} and \ref{fig14}, 
are the production
rates of direct photons (solid line) and $\pi^0$'s (dashed and
dot-dashed lines) for central $Au+Au$ collisions at the RHIC and
LHC energies. The rate of $\pi^0$ production is calculated with the same
jet fragmentation functions employed in this paper and convoluted
with jet production cross sections \cite{wang2}.
We can see that without jet quenching,
$\pi^0$ production rate is about 20 times larger than the
direct photons at $p_T=10$ GeV/$c$ at $\sqrt{s}=200$ GeV. 
{}Fortunately, jet quenching 
due to parton energy loss can significantly reduce $\pi^0$
rate at large $p_T$ as shown by the dot-dashed line. However,
one still has to face $\pi^0$'s about 3 times higher than
the direct photons at $p_T=10$ GeV/$c$. At larger $p_T$,
the situation improves, but one loses the production rate.
Since the isolation cut method normally employed in $pp$ collisions
to reduce the background for direct photons does not work any more, the
only way one can identify them with high accuracies has to be through 
the means of improved detectors.

Similarly, we also list in Table~\ref{tab2} the number of
$\gamma+{\rm jet}$ events per year per unit rapidity and
unit (GeV) $E_T$ at the LHC energy. We assume
a luminosity of ${\cal L}=2\times 10^{27}$
cm$^{-2}$s$^{-1}$ with 50 operation days per year for
$Au+Au$ collisions. The production rates are
reasonably high due to both the high luminosity and collider
energy. However, the corresponding background of $\pi^0$'s
is also high (see Fig.~\ref{fig14}) which may make 
the detection of direct photons more difficult.

At the LHC energy, $\sqrt{s}=5.5$ TeV, the production rate
for $Z^0+{\rm jet}$ becomes large even for reasonably large $P_T^{Z^0}$.
Listed in Table~\ref{tab3} are the number of $Z^0+{\rm jet}$ events
per year per unit rapidity integrated over $P_T^{Z^0}$ with
different low cut-off values. The production cross sections
are provided by T. Han based on calculations as described in
Ref.~\cite{han}. Note that the given $Z^0$ production rates are
integrated ones, thus appearing to be larger than the differential
rate of direct photon production at the same transverse momentum.
One can detect $Z^0$ through the dilepton channel which has almost no
background in the range of the dilepton invariant mass near $M_{Z^0}$.
One can then apply the same procedure as we have discussed in this paper
for direct photon events and measure the modification of the effective
jet fragmentation function due to parton energy loss. However,
the drawback of using the dilepton channel of $Z^0$ decay is that
the effective number of events via this channel is about 6.7\% of the
total number of $Z^0$ events.

\begin{center}
\begin{table}

\begin{tabular}{|l|llll|}
$E_T^{\gamma}$ (GeV) & 7 & 10 & 15 & 20 \\ \hline
$dN^{\gamma-jet}/dydE_T/$year & 20500 & 3550 & 400 & 70 \\
\end{tabular}
\bigskip
\caption{Rate of direct photon production in central $Au+Au$
  collisions at $\protect\sqrt{s}=200$ GeV, with luminosity 
  ${\cal L}=2\times 10^{26}$ cm$^{-2}$s$^{-1}$ and 100 operation
  days per year.}
\label{tab1}
\end{table}
\end{center}

\vspace{-0.5in}

\begin{center}
\begin{table}
\begin{tabular}{|l|lll|}
$E_T^{\gamma}$ (GeV) & 40 & 50 & 60 \\ \hline
$dN^{\gamma-jet}/dydE_T/$year & 2880 & 1070 & 490 \\
\end{tabular}
\bigskip
\caption{Rate of direct photon production in central $Au+Au$
  collisions at $\protect\sqrt{s}=5.5$ TeV, with luminosity 
  ${\cal L}=2\times 10^{27}$ cm$^{-2}$s$^{-1}$ and 50 operation
  days per year.}
\label{tab2}
\end{table}
\end{center}

\vspace{-0.5in}

\begin{center}
\begin{table}

\begin{tabular}{|l|lll|}
$P_T^{Z^0}$ (GeV) & $>20$ & $>40$ & $>60$ \\ \hline
$dN^{Z^0-jet}/dy/$year & 21100 & 7700 & 3470 \\
\end{tabular}
\bigskip
\caption{Rate of $Z^0$ production in central $Au+Au$
  collisions at $\protect\sqrt{s}=5.5$ TeV, with luminosity 
  ${\cal L}=2\times 10^{27}$ cm$^{-2}$s$^{-1}$ and 50 operation
  days per year.}
\label{tab3}
\end{table}
\end{center}

\section{Conclusions}

In summary, we have studied systematically how one can measure the
parton energy loss in $\gamma+{\rm jet}$ events in central high-energy
heavy-ion collisions within the framework of a modified fragmentation
function model. We have demonstrated that an effective fragmentation
function of the photon-tagged jets can be extracted from the inclusive
charged hadron spectrum in the opposite direction of the tagged photon.
We have estimated the background from large $p_T$ hadron production to
be small as compared to hadron production from the jet fragmentation
for relatively large values of $p_T$. We also provided estimates
of the lower limits on $E_T^\gamma$ in central $A+A$ collisions
in order for such extraction to be possible.
We further show that the effective fragmentation function is
sensitive to the parton energy loss possibly experienced by
the parton during its propagation through the produced dense
matter. The sensitivity is characterized by the so-called modification
factor via the comparison of the effective fragmentation function
in $AA$ with the one in $pp$ collisions. 

We have explicitly taken into account the $E_T$-smearing of the 
photon-tagged jets for a fixed value of $E_T^\gamma$ due to 
initial state radiations. We also demonstrated
that the defined modification factors in $pA$ collisions probe
the $E_T$ broadening due to multiple initial and final state 
parton scatterings with the beam nucleons. One should subtract out
the effect of such $E_T$ broadening when extracting 
the parton energy loss from the modification factor in $AA$ collisions.
We have also made detailed analysis of the modification factor
within our model and studied the sensitivity to different
forms of the parton energy loss, {\it e.g.}, $A$ dependence
and the effective jet profile as a function of the azimuthal
angle in the transverse plane. We have also applied
our model to jet quenching in deeply inelastic lepton-nucleus
collisions from which one can extract the parton energy
loss inside a cold nuclear matter. Finally, we have examined
the experimental feasibilities of our proposed study. Our estimates
show that in order to have accurate measurements, one need
somewhat increased luminosity about 
${\cal L}=2\times 10^{27}$ cm$^{-2}$s$^{-1}$ at the RHIC energy.
At the LHC energy, one can alternatively use $Z^0$ as the trigger
and study the associated jet fragmentation.

\nopagebreak
\section*{Acknowledgements}
We thank I.~Sarcevic for helpful discussions and her early
collaboration. We also thank T. Han for providing the cross sections 
of $Z^0$ production and helpful discussions. X.-N. W. would like to 
thank J.~B.~Carroll and D.~Kharzeev for helpful discussions. 
This work was supported by the Director, Office of Energy 
Research, Division of Nuclear Physics of the Office of High 
Energy and Nuclear Physics of the U.S. Department of Energy 
under Contract Nos. DE-AC03-76SF00098 and DE-FG03-93ER40792.


\begin{references}
\bibitem{aco}D. A. Appel, Phys. Rev. D {\bf 33}, 717 (1986);
        J. P. Blaizot and L. D. McLerran, Phys. Rev. D {\bf 34}, 2739 (1986);
        M. Rammerstorfer and U. Heinz, Phys. Rev. D {\bf 41}, 306 (1990);
        S. Gupta, Phys. Lett. {\bf B347}, 381 (1995).
\bibitem{qn1}M. Gyulassy and M. Pl\"umer, Phys. Lett. {\bf B243}, 432 (1990);
        M. Thoma and M. Gyulassy, Nucl. Phys. {\bf B351}, 491 (1991);
        M. Pl\"umer, M. Gyulassy and X.-N. Wang, Nucl. Phys. {\bf A590},
        511c (1995).
\bibitem{lpm1}M. Gyulassy and X.-N. Wang, Nucl. Phys. {\bf B420}, 583 (1994);
        X.-N. Wang, M. Gyulassy and M. Pl\"umer, Phys. Rev. D {\bf 51},
        3436 (1995).
\bibitem{lpm2}R. Baier, Yu.~L.~Dokshitzer, S.~Peign\'e, D.~Schiff, 
        Phys. Lett. {\bf B345}, 277 (1995); R. Baier, Yu. L. Dokshitzer, 
        A. Mueller, S. Peign\'e and D. Schiff, Nucl. Phys. {\bf B478},
        577 (1996); hep-ph/9607355; hep-ph/9608322.
\bibitem{lpm0}L. D. Landau and I. Ya. Pomeranchuk, Dokl. Akad. Nauk
        SSSR {\bf 92}, 535, 735 (1953); A.~B.~Midgal, Phys. Rev.
        {\bf 103}, 1811 (1956).
\bibitem{wg90}X.-N. Wang and M. Gyulassy, proceedings of the
         Fourth Workshop on Experiments and Detectors for RHIC,
         July 2-7, 1990, Brookhaven National Laboratory, 
         Eds. M. Fatyga and B. Moskowitz, p.79, BNL-52262.
\bibitem{qn2}X.-N. Wang and M. Gyulassy, Phys. Rev. Lett. 
        {\bf 68}, 1480 (1992).
\bibitem{wang2}X.-N. Wang, in preparation.
\bibitem{whs}X.-N. Wang, Z. Huang and I. Sarcevic, Phys. Rev. Lett. 
        {\bf 77}, 231 (1996).
\bibitem{report}X.-N. Wang, hep-ph/9605214, Phys. Rept. in press;
        X.-N. Wang, in {\it Quark-Gluon Plasma II}, R. C. Hwa (ed.)
        (World Scientific, 1995).
\bibitem{mattig}P. M\"{a}ttig, Phys. Rep. {\bf 177}, 141 (1989).
\bibitem{bkk}J. Binnewies, B. A. Kniehl and G. Kramer, Z. Phys.
        C{\bf 65}, 471 (1995).
\bibitem{hijing}X.-N. Wang and M. Gyulassy, Phys. Rev. D {\bf 44}, 3501 (1991);
        Comp. Phys. Comm. {\bf 83}, 307 (1994).
\bibitem{soper}S. D. Ellis, Z. Kunszt and D. E. Soper, Phys. Rev. Lett. 
        {\bf 62}, 726 (1989); Phys. Rev. D {\bf 40}, 2188 (1989);
        Phys. Rev. Lett. {\bf 69}, 1496 (1992).
\bibitem{pythia} T.~Sj\"{o}strand and M.~van Zijl, Phys. Rev. D {\bf 36},
        2019 (1987); 
        T.~Sj\"{o}strand, Comput. Phys. Commun. {\bf 39}, 347 (1986);
        T.~Sj\"{o}strand and M.~Bengtsson, {\em ibid.} {\bf 43}, 367 (1987).
\bibitem{ua1}UA1 Collab., G. Arnison {\it et al.}, Phys. Lett. {\bf B 172},
        461 (1986); C. Albajar {et al.}, Nucl. Phys. {\bf B309}, 405 (1988).
\bibitem{mrs}A. D. Martin, W. J. Stirling and R. G. Roberts, Phys. Lett.
        {\bf B306}, 145 (1993).
\bibitem{cronin1}P. Bordalo {\it et al.}, Phys. Lett. {\bf B193}, 373 (1987).
\bibitem{cronin3}E609 Collaboration, M.~D.~Corcoran, {\it et al.},
        Phys. Lett. {\bf B259}, 209 (1991).
\bibitem{cronin2}D.~M.~Alde {\it et al.}, Phys. Rev. Lett. {\bf 64}, 
        2479 (1990).
\bibitem{gv1}S.Gavin and M. Gyulassy, Phys. Lett. {\bf B214}, 241 (1988);
         S. Gavin and R. Vogt,hep-ph/9610432.
\bibitem{gm}S. Gavin and J. Milana, Phys. Rev. Lett. {\bf 68},
        1834 (1992). 
\bibitem{NAG}G. Nilsson, B. Andersson and G. Gustafson, Phys. Lett.
        {\bf B83}, 379 (1979).
\bibitem{niko}N. N. Nikolaev, Z. Phys. {\bf C5}, 291 (1980).
\bibitem{bialas}A. Bialas and M. Czyzewski, Phys. Lett. {\bf B222},
        132 (1989)
\bibitem{gp_eA}M. Gyulassy and M. Pl\"umer, Nucl. Phys. {\bf B346},1 (1990).
\bibitem{eskola}K. J. Eskola, Nucl. Phys.{\bf B400}, 240 (1993).
\bibitem{e665}P. Madden, Ph. D. thesis, Univ. of California, San Diego,1996.
\bibitem{hp1} {\it Hard Processes in Hadronic Interactions},
        Eds. H. Satz and X.-N. Wang, Int. J. Mod. Phys. {\bf A10}, 2881-3090
        (1995).
\bibitem{han}V. Barger, T. Han, J. Ohnemus and D. Zeppenfeld,
         Phys. Rev. Lett. 62, 1971 (1989); V.~Barger, T.~Han, 
         J.~Ohnemus and D.~Zeppenfeld, 
         Phys. Rev. D40, 2888 (1989); and references therein.

\end{references}
\end{document}